\documentclass[aps,epsf,twocolumn,showpacs]{revtex4-1}
\usepackage{amsmath}
\usepackage{epsfig}

\begin{document}

\title{Critical Behavior of the Three-Dimensional Ising model with Anisotropic Bond Randomness at the Ferromagnetic-Paramagnetic Transition Line}

\author{T. Papakonstantinou$^1$}

\author{A. Malakis$^1$}

\affiliation{$^1$Department of Physics, Section of Solid State
Physics, University of Athens, Panepistimiopolis, GR 15784
Zografou, Athens, Greece}

\date{\today}

\begin{abstract}
We study the $\pm J$ three-dimensional Ising model with a
spatially uniaxially anisotropic bond randomness on the simple
cubic lattice. The $\pm J$ random exchange is applied in the $xy$
planes, whereas in the z direction only a ferromagnetic exchange
is used. After sketching the phase diagram and comparing it with
the corresponding isotropic case, the system is studied, at the
ferromagnetic-paramagnetic transition line, using parallel
tempering and a convenient concentration of antiferromagnetic
bonds ($p_z=0 ; p_{xy}=0.176$). The numerical data point out
clearly to a second-order ferromagnetic-paramagnetic phase
transition belonging in the same universality class with the 3d
random Ising model. The smooth finite-size behavior of the
effective exponents describing the peaks of the logarithmic
derivatives of the order parameter provides an accurate estimate
of the critical exponent $1/\nu=1.463(3)$ and a collapse analysis
of magnetization data gives an estimate $\beta/ \nu=0.516(7)$.
These results, are in agreement with previous studies and in
particular with those of the isotropic $\pm J$ three-dimensional
Ising at the ferromagnetic-paramagnetic transition line,
indicating the irrelevance of the introduced anisotropy.
\end{abstract}

 \maketitle

\section{Introduction}
\label{sec:1}

Ising spin glass models yield phase diagrams with distinctively
complex ordered phases, in $d=3$. These models, although
relatively simple in their formulation, have been proposed to
describe complex systems exhibiting frustration, e.g., materials
such as ${\mathrm{Fe_{1-xi}M n_{xi}TiO_3}}$ and ${\mathrm{Eu_{1-x}
Ba_x MnO_3}}$~\cite{glass-metal,glass-metal2,glass-metal3}, neural
networks~\cite{hopfield1} etc.

The simplest of such models, but most influential over the years,
is the Edwards Anderson model~\cite{ea-model,BindY86} defined by
the Hamiltonian
\begin{equation}
\label{eq:1}
H=-\sum_{\langle ij \rangle}J_{ij}s_{i}s_{j},
\end{equation}
where the summation is over nearest-neighbors, $s=\pm 1$ and
$J_{ij}$ denotes the uncorrelated quenched exchange interaction.
There are two popular quenched random disorder distributions, the
Gaussian distribution of random bonds with zero mean and unity
standard deviation and the bimodal distribution of $J_{ij}$ given
by
\begin{equation}
\label{eq:2} P(J_{ij}) = p\delta(J_{ij}+1)+(1-p)\delta(J_{ij}-1)
\end{equation}

Recently, the spatially uniaxially anisotropic $d=3$ spin glass
system has been solved exactly on a hierarchical lattice by Guven
et al.~\cite{anis_1}. Their general study revealed a rich phase
diagram topology and several new interesting features. The
Hamiltonian of this anisotropic variant, differentiates, along $z$
axis, the probability distribution of quenched randomness, but
also the strength of the exchange interaction and can be written
\begin{equation}
\label{eq:3} H=-\sum_u\sum_{\langle ij
\rangle_u}J_{ij}^us_{i}s_{j},
\end{equation}

Accordingly, the bimodal distribution of $J_{ij}^u$ takes the more
general form
\begin{equation}
\label{eq:4} P(J_{ij}^u) =
p_u\delta(J_{ij}^u+J^u)+(1-p_u)\delta(J_{ij}^u-J_{ij}^u)
\end{equation}
where $u$ denotes the $z$ axis $(u=z)$ or the $xy$ planes
$(u=xy)$, $J_{ij}^u$ denote the corresponding exchange
interactions and $p_u$ are the probabilities of two neighboring
spins ($ij$) having antiferromagnetic interaction along $z$ axis
or the $xy$ planes.

\begin{figure}[htbp]
\includegraphics*[width=8.5 cm]{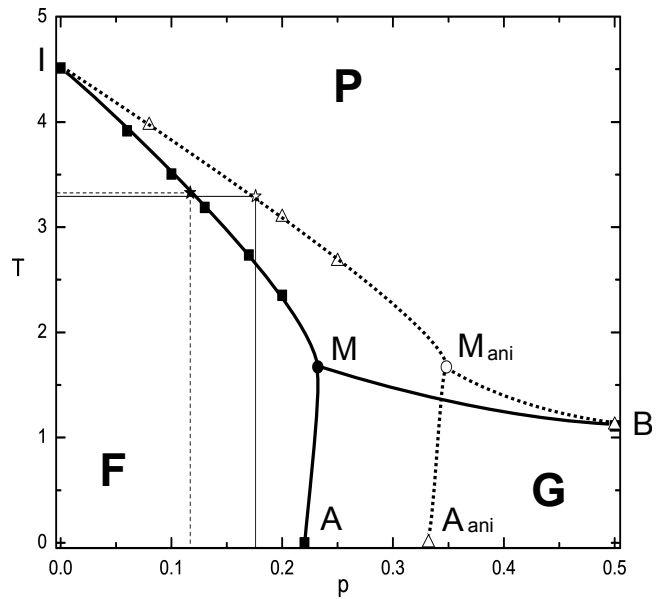}
\caption{\label{fig:1}(color on line) Phase diagram points (full
squares, asterisk, and dot) of the isotropic Edwards-Anderson
model are taken
from~\cite{hasen_fp,hasen_mcp,katz_pg,hartmann_sggs}. The phase
diagram is drawn in solid lines and separates the three phases:
ferromagnetic (F), spin-glass (G), and paramagnetic (P). These
lines meet at the multicritical full dot point (M). Dash lines
(and open symbols) illustrate the phase diagram of the anisotropic
model $p_z=0, p=p_{xy}$. The open triangles, asterisk and dot
points are discussed in more detail in the text. In particular,
the full asterisk ($p=0.117$) corresponds to the improved model
case of~\cite{hasen_fp} and the open asterisk ($p=p_{xy}=0.176$)
to the improved case studied here.}
\end{figure}

The standard isotropic case, defined by Eq.~(\ref{eq:1}) and
Eq.~(\ref{eq:2}), corresponds to $J_{ij}^z=J_{ij}^{xy}$ and
$p_z=p_{xy}$. The global phase diagram of this isotropic case, is
sketched in Fig.~\ref{fig:1} and, as shown, there exist (in $d=3$)
three distinctive phases, ferromagnetic, paramagnetic and glassy
phases. All transitions among these phases, are believed to be of
second order and also to belong to different universality classes.
Several accurate studies have been carried out, to determine the
critical behaviors along these transition lines, for the
finite-temperature phase transitions
\cite{FG_1,hasen_fp,hasen_sg,katz_pg,jorg,campb_sg,balle_sg,pala_sg,kawa_is,bill-11}.
Most of these studies concern the ferromagnetic-paramagnetic (FP)
and the glassy-paramagnetic (GP) lines. There is also a recent
study for the ferromagnetic-glassy transition line~\cite{FG_1}.
The transition lines meet in a multicritical
point~\cite{hasen_mcp,mcp_2d,ozeki-ito,mcp_2,doussal_harris_1,doussal_harris_2}
located along the Nishimory line
~\cite{nishimori_book,nishimori_1980,nishimori_1981,nishimori_1986}
with coordinates $T_{M}=1.6692(3), p_{M}=0.23180(4)$
\cite{hasen_mcp,mcp_2d,ozeki-ito,mcp_2}. The FP transition line,
starts at the pure Ising model ($p=0$), for which all critical
properties have been extensively studied and a recent very
accurate estimate of the correlation length exponent is
$\nu=0.63002(10)$~\cite{Hase10,camp_is,guida_is,Butera-02} with a
critical temperature $T_c=4.5115232(16)$
~\cite{Hase10,camp_is,guida_is,Butera-02,deng_is,blote_is}. As
shown by Hasenbusch et al.~\cite{hasen_fp}, the introduction of
the $\pm J$ quenched randomness changes the universality of the
model to that of the random Ising model (RIM) or randomly diluted
Ising model (RDIs)~\cite{kawa_is,hukushima_fp,hasen_fp}, in which
several spin models appear to belong. These, include models such
as the randomly site and bond diluted, the random bond
~\cite{balles_rdi,hasen_rdi,pelis_rdi,berche-04,theod_fytas}, the
random bond Blume-Capel~\cite{mal_3drbc} model and, of course, the
already mentioned isotropic $\pm J$ three-dimensional Ising at the
ferromagnetic-paramagnetic transition line~\cite{hasen_fp}. An
accurate estimate of the correlation length exponent for these
$d=3$ FP transitions, characterizing the RIM universality class,
is that given by ~\cite{balles_rdi} $\nu=0.6837(53)$.

For the glassy-paramagnetic (GP) transition line ($MB$ line in
~Fig.\ref{fig:1}), most of the work has been curried out at
$p=\frac{1}{2}$ ~\cite{ogiel-85,ogiel-85-2,sing_sg,bhatt_sg,
kawa_young,bern_sg,berg_sg,pala_sg,mari_sg,balle_sg,mari_sg-2,
mari_sg-3,naka_sg,pleim_sg,campb_sg,campb_sg-2,katz_pg}. However,
because of severe inherent difficulties, due to both strong
frustration and disorder effects in this region, there is a large
spread in the estimates of critical exponents and there remain
some questions related to universality. This is very clearly
reflected in Table I of Katzgraber et al.~\cite{katz_pg} in which
one can observe a very large spread in the correlation length
exponent, and even in the estimates of critical temperature. We
shall quote from this paper, the estimates $\nu=2.39(5)$ and
$T_c=1.120(4)$, which apply to the present isotropic $\pm J$
spin-glass model. We will also quote the more recent estimates
$\nu=2.45(15)$, $T_c=1.109(10)$~\cite{hasen_sg}. Finally, from the
recent work by Ceccarelli et al.~\cite{FG_1}, we know that the
correlation length exponent for the ferromagnetic-glassy (FG)
transitions has been estimated to be $\nu=0.96(2)$.

Several other features of the critical behaviors and the global
phase diagram are known for the isotropic case. Ground state
calculations by Hartmann~\cite{hartmann_sggs} ($p_A=0.222(5)$),
indicated a reentrant FG transition line. This was nicely verified
by the finite temperature study of Ceccarelli et al.~\cite{FG_1},
since they estimated $p(T=0.5)=0.2271(2)$, predicting the
ferromagnetic-glassy transition line to be slightly reentrant.
These results are in accordance with the Nishimori
expectations~\cite{nishimori_1980,nishimori_1981} that, this line
cannot be forward and are also reflected in renormalization-group
calculations~\cite{miglio_1,berker-05,anis_1}.

In this paper we will focus on the nature of the FP transition for
the anisotropic model described by Eq.~(\ref{eq:3}) and
Eq.~(\ref{eq:4}) by considering the particular model case $p_z=0 ;
p_{xy}\leq 0.5$ with $J_{ij}^z=J_{ij}^{xy}$. The present study is
part of a research program to study by Monte Carlo methods the
general spatially uniaxially anisotropic $d=3$ spin glass system
considered by Guven et al.~\cite{anis_1}. The main motivation is
to provide numerical evidence for the universality question, by
investigating possible effects caused, by the introduced
anisotropy, on the critical exponents along the three different
transition lines. We hope that our attempts may also play some
role in future studies and will provide new information for the
critical behavior of frustrated systems.

We start  by describing our current perception on the global phase
diagram of the above anisotropic case. This phase diagram is
sketched also in Fig.~\ref{fig:1}, by the dash lines. As shown,
there exist again(in $d=3$) three distinctive phases,
ferromagnetic, paramagnetic and glassy phases. The points shown on
the FP transition line have been located from the peaks of the
magnetic sample-averaged susceptibility for a lattice of size
$L=16$. The open asterisk (see also the solid drop lines) on this
line indicate a special point corresponding to $p_{xy}^*=0.176$.
This choice was inspired by the improved model case analyzed by
Hasenbusch et al.~\cite{hasen_fp} for the isotropic model (see on
the isotopic phase diagram the case with $p^*=0.117$ and the
corresponding dash drop lines). According, to Hasenbusch et
al.~\cite{hasen_fp} for the isotropic case at $p^*=0.117(3)$ the
leading scaling corrections vanish and this case provides an
improved model for accurate exponent estimation. Our quest for a
convenient and improved model case for the anisotropic FP
transition, directed as to find the phase diagram point
corresponding to about the same temperature as the isotropic phase
diagram point. As can be seen, by comparing the drop lines, this
would suggest the case $p_{xy}^*=0.176$. Alternatively, let us try
to satisfy the relation $(1-p_{xy}^*(T))=1.5(1-p^*(T))-0.5$
between the two FP transition lines. This prognostic method
assumes that phase diagram points corresponding to the same
temperature would have (approximately) the same ratio of
ferromagnetic to antiferromagnetic interactions. Putting
$p^*=0.117(3)$ in this relation, we find $p_{xy}^*=0.1755(30)$,
which is practically the above value for the anisotropic case. As
it will be seen in the sequel, the proposed case $p_{xy}^*=0.176$
provides indeed an excellent case and gives, for the anisotropic
model, a very accurate estimate of the correlation length
exponent($\nu=0.6835(25)$).

The rest of the anisotropic phase diagram (see also the caption of
Fig.~\ref{fig:1}), was mainly constructed on the premise that the
above relation may hold approximately for the lines separating the
ferromagnetic phase from the other two phases. Thus, applying this
prognostic to the point $p(T=0)=p_A=0.222(5)$ of the isotropic
case, we find that $p_{xy}(T=0)=p_{A_{ani}}=0.333(5)$. Quite
remarkably we found from our ground state calculations a value
that perfectly verify this prediction. Based on the rather short
calculation of ground states on lattices sizes $L=6,8,10$ we
found, by a collapse method, $p_{A_{ani}}=0.332(12)$ astonishingly
close to the predicted value. After this, the multicritical point
for the anisotropic case, shown in Fig.~\ref{fig:1} has been
placed by a mere application of the above relation giving
$p_{M_{ani}}=0.34770(4)$. One should, of course, realize that this
prognostic is only an approximation, which could not be expected
to apply in the more general case. It is possible that, it is only
successful for the present case giving a reasonable approximation
for the FP and FG transition lines. To complete our perception of
the anisotropic phase diagram, we point out that for the case
$p_z=0 ; p_{xy}\leq 0.5$ we found from Monte Carlo simulations a
phase diagram point which appears to coincide with the isotopic
phase diagram point $B$. For instance, extrapolating the
$L=6,8,10$ Binder's fourth order cumulant crossings of the overlap
order parameter we found $T_{B_{ani}}=1.07(4)$, very close to the
corresponding estimate of the isotropic case. This finding appears
to be a very interesting and physically appealing prediction. A
proof that $T_{B_{iso}}=T_{B_{ani}}$ exactly, may also help a
better understanding of the general universality question. In
closing this discussion, we shall mention that the introduction of
anisotropy changes in general the phase diagram and possibly the
symmetry of multicritical points, as discussed by Guven et
al.~\cite{anis_1}. These multicritical points, such as in the
present model case ($p_z=0 ; p_{xy}\leq 0.5$), do not necessarily
obey Nishimori conditions and one can not exclude the possibility
of a ferromagnetic-glassy transition line being forward. However,
our prognostic method, insist that the present anisotropic FG
transition line is reentrant.

The rest of the paper is laid out as follows: In the following
subsection we give a description of our numerical approach
utilized to derive numerical data for large ensembles of
realizations of the disorder distribution and lattices with linear
sizes within the range $L\in \{8-44\}$. Then in
subsection~\ref{sec:2b} the finite-size scaling (FSS) scheme is
described in some detail. In Sec.~\ref{sec:3} we present all our
FSS attempts, that give good estimates of all critical exponents
and verify that the present anisotropic model belongs to the
universality class of RIM. Our conclusions are summarized in
Sec.~\ref{sec:4}.

\section{MONTE CARLO SIMULATIONS AND FINITE-SIZE SCALING SCHEME}
\label{sec:2}
\subsection{MONTE CARLO METHOD}
\label{sec:2A} In the present paper, we shall use our recent
approach to disordered systems~\cite{mal_3drbc}, based on a
parallel tempering (PT) practice. Our PT protocol will use an
adequate number of Metropolis~\cite{metro53} sweeps of the
lattice, as an elementary Monte Carlo step, so that the
correlation times of the PT protocol will be, for all temperatures
used, very close to $1.0$. We mention here, that parallel
tempering, combined with Metropolis algorithm has been used also
by Ceccarelli et al.~\cite{FG_1} in their study of the FG
transition. Further, as pointed out by Hasenbusch et
al.~\cite{hasen_fp}, in their study of the isotropic case, the
Metropolis algorithm may be more effective than cluster dynamics,
for intermediate lattice sizes, as a result of frustration effects
present in the $\pm J$ models.

Our PT protocol (also our FSS approach) has
been described in detail in our recent
paper~\cite{mal_3drbc,malakis12} and only a brief summary is given
here. For the estimation of the critical properties we generate MC
data that cover several finite-size anomalies of the finite
systems of linear size $L$. The PT approach is carried over to a
certain temperature range depending on the lattice size. These
temperatures are selected in such a way that the exchange rate is
$0.5$, using a practice similar to that suggested in
Ref.~\cite{bittner11}. The appropriate temperature sequences were
generated via short preliminary runs in which we apply a simple
histogram method~\cite{Swendsen87,Newman99,LandBind00} to determine from the
energy probability density functions the temperatures, satisfying
the above exchange condition~\cite{bittner11,mal_3drbc}. The
preliminary runs cover several disorder realizations and the
average over the temperatures sequences provides us with a
protocol with very small variation in the  exchange rate
condition, as one moves from one realization to the other.

The proposed MC scheme was carefully tested for all lattice sizes
before its implementation for the generation of MC data. The tests
included the estimation of the MC times necessary for
equilibration and thermal averaging process applied to a
particular disorder realization, and involved also the observation
of running sample-averages of several thermodynamic quantities,
such as the magnetic susceptibility, at a temperature close to
criticality. Provided we use reasonably long MC times for
equilibration and thermal averaging we find an effective
cancellation of statistical errors from the sample-averaging
process, since the sample-to-sample fluctuations are larger (by an
order of magnitude) than the usual statistical errors (in the
thermal averaging process).

The present simulation task is, of course, quite considerable,
since we have to satisfy both a good equilibration of the system
and also sum over a very large number of disorder realizations.
The number of realizations influence the accuracy of our data, and
the proper selection of temperatures influence their suitability
for the locations of the finite-size anomalies. Since we are
implementing a PT approach, based on temperatures corresponding to
an exchange rate $0.5$, we are selecting proper temperature
sequences consisting of a number of (say 3 or 5) different
temperatures and averaging for this set of temperatures over a
relatively large number ($\sim 1000$) of disorder realizations.
This yields a number of (3 or 5) points of the averaged curves
$[Z]_{av}$, where $Z$ denote the thermal average of some
thermodynamic quantity, such as the magnetic susceptibility. The
PT protocol is repeated several times (depending on the linear
size $L$) by using new sets of temperatures(translated with
respect to the previous sets) and a final dense set of points is
obtained. The above is an efficient and most importantly quite
accurate practice, since we finally obtain averaged curves
corresponding to a very large number of realizations, describing
with high accuracy all the averaged finite-size anomalies of the
system. Monte carlo data were collected for systems of linear
sizes $L={8,12,16,20,24,28,32,36,40,44}$ and the FSS analysis was
performed to the averaged quantities obtained from these MC data.

\subsection{FINITE-SIZE SCALING SCHEME}
\label{sec:2b}

In the standard approach of FSS for a random system, a large
number of disorder realizations has to be used in the summations
in order to obtain good sample-averages of the basic thermodynamic
quantities $Z$, which are the usual thermal averages of a single
disorder realization. From the disorder averages $[Z]_{av}$ we
obtain their finite-size anomalies, denoted here as
$[Z]_{av}^{\ast}$. These finite-size anomalies will be used in our
FSS attempts, following a quite common practice~\cite{chate-01}.
Their temperature locations, denoted by $T_{[Z]_{av}^{\ast}}$ will
be used in the sequel in our FSS attempts. Thus, our study
concerns the critical exponents describing the disorder-averaged
behavior and we do not attempt a FSS analysis based on sample
dependent pseudocritical temperatures. The later is, a more
demanding alternative approach~\cite{wiseman-98,DSFish95,Chay86},
which considers the individual sample dependent maxima (anomalies)
and the corresponding sample dependent pseudocritical
temperatures. Note that, for disordered systems one could make in
principle a clear distinction between typical and averaged
exponents~\cite{DSFish95,Chay86}.

From the MC data, several pseudocritical temperatures may be
estimated, corresponding to finite-size anomalies and these are
expected to follow a power-law shift behavior
$T_{[Z]_{av}^{\ast}}=T_{c}+b_{Z}\cdot L^{-1/\nu}$. The
traditionally used specific heat and magnetic susceptibility
peaks, as well as the peaks corresponding to the following
logarithmic derivatives of the powers $n=1$, $2$, and $n=4$ of the
order parameter with respect to the inverse temperature
$K=1/T$~\cite{ferrenberg91},
\begin{equation}
\label{eq:5} \frac{\partial \ln \langle M^{n}\rangle}{\partial
K}=\frac{\langle M^{n}H\rangle}{\langle M^{n}\rangle}-\langle
H\rangle,
\end{equation}
and the peak corresponding to the absolute order-parameter
derivative
\begin{equation}
\label{eq:6} \frac{\partial \langle |M|\rangle}{\partial
K}=\langle |M|H\rangle-\langle |M|\rangle\langle H\rangle,
\end{equation}
will be located and used in our fitting attempts.

The behavior of the maxima of the logarithmic derivatives of the
powers $n=1$, $2$, and $n=4$ of the order parameter with respect
to the inverse temperature, which as is well known scale as $\sim
L^{1/\nu}$ with the system size~\cite{ferrenberg91}, will be seen
to provide a smooth root for the estimation of the correlation
length exponent $\nu$. Once the exponent $\nu$ is well estimated,
the behavior of the values of the peaks corresponding to the
absolute order-parameter derivative, which scale as $\sim
L^{(1-\beta)/\nu}$ with the system size~\cite{ferrenberg91}, gives
one route for the estimation of the magnetic exponent ratio
$\beta/\nu$.

For the estimation of the critical temperature, we shall use
mainly a simultaneous fitting approach of the several
pseudocritical temperatures mentioned above. However, from the MC
data for the disorder averaged magnetization, it is possible to
follow an optimum collapse method which will provide
simultaneously estimates for the critical exponents $\beta/ \nu ,
{1/ \nu}$ as well as the critical temperature $T_c$. From the
scaling hypothesis
\begin{equation}
\label{eq:7} [\langle |M|\rangle]_{av}=M(T,L)=L^{-\beta/
\nu}{f}[(T-T_c)L^{1/ \nu}],
\end{equation}
and the disorder averaged magnetization data, we attempt in the
next section the estimation of critical behavior by using a
recently published collapse method that makes use of the downhill
simplex algorithm~\cite{collapse}.

\section{ESTIMATION OF CRITICAL BEHAVIOR}
\label{sec:3}

\begin{figure}[htbp]
\includegraphics*[width=9.5 cm]{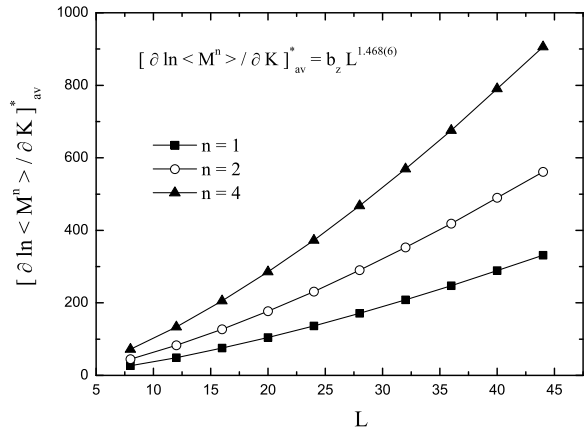}
\caption{\label{fig:2} (color online) FSS behavior of the peaks of
the logarithmic derivatives of the powers $n=1$, $2$, and $n=4$ of
the order parameter with respect to the inverse temperature. The
estimate for the exponent $1/\nu=1.468(6)$ (shown in the panel) is
obtained by applying a simultaneous fitting attempt to a simple
power law in the whole size range $L=8-44$.}
\end{figure}

\begin{figure}[htbp]
\includegraphics*[width=9.5 cm]{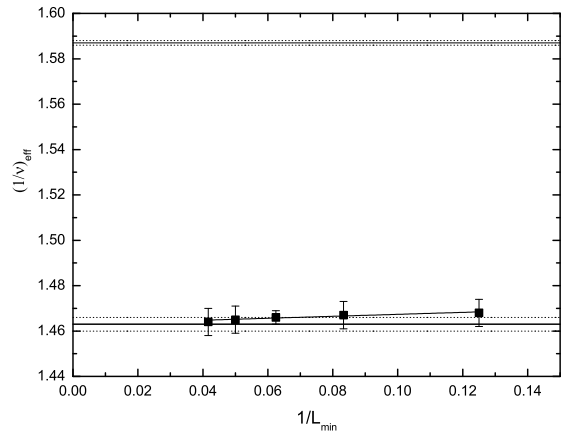}
\caption{\label{fig:3} (color online) Illustration of the behavior
of the effective exponents $(1/\nu)_{eff}$. The solid line drawn
in the panel, together with the dot lines, indicate the critical
exponent range for the present model. The linear behavior
illustrated give the accurate estimation $1/\nu=1.463(3)$. For
comparison the analogous narrow range $1/\nu=1.587(1)$
\cite{Hase10}, for the pure 3d Ising model is shown in the upper
part of the panel.}
\end{figure}

As pointed out earlier, the present study uses the finite-size
anomalies of the sample average of the logarithmic derivatives of
the powers ($n=1,2,4$) of the order parameter with respect to the
inverse temperature, for the estimation of the correlation length
exponent $\nu$. It will be seen that, the finite size scaling
behavior of the corresponding peaks provides for the present model
an attractive and smooth approach to this estimation. Assuming
that these finite-size anomalies ($[Z]_{av}^{\ast}$) of the
disorder averages $[Z]_{av}$, where $Z$ is the thermal average
given by Eq.~(\ref{eq:5}), scale as $\sim L^{1/\nu}$ with the
system size~\cite{ferrenberg91} we estimate in figures
~\ref{fig:2} and ~\ref{fig:3} this exponent. A very good scaling
behavior is observed already by using the whole size range of our
Monte carlo data $L=[8-44]$. The simultaneous fitting attempt to
the expected power-law behavior, illustrated in Fig.~\ref{fig:2},
gives the estimate shown in the panel: ${1}/\nu=1.468(6)$.

By varying the $L_{min}$, in these simultaneous fitting attempts,
we obtain a sequence of effective exponents depending on the
minimum size used. The behavior of these effective exponents is
shown in Fig.~\ref{fig:3}. The smooth, almost perfectly linear,
behavior of these effective exponents enables us to estimate
confinedly ${1}/\nu=1.463(3)$. The error range of this estimation
is indicated in Fig.~\ref{fig:3} by the dot lines and is compared
with the corresponding 3d pure Ising model for which an extremely
accurate estimation is available\cite{Hase10}. The present
estimate for the correlation length exponent $\nu=0.6835(25)$,
compares well with the estimate $\nu=0.683(3)$ of Hasenbusch et
al.~\cite{hasen_fp}, for the corresponding isotropic $\pm J$ Ising
model at the ferromagnetic-paramagnetic transition line and is in
excellent agreement with the estimate $\nu=0.6837(53)$, of the
extensive numerical investigations of Ballesteros \emph{et
al.}~\cite{balles_rdi} for the site-diluted Ising model. The good
behavior of the effective exponents in Fig.~\ref{fig:3}, giving
improved error bounds, suggest the present model as a suitable
candidate for further attempts and refinements of this exponent,
possibly following the more sophisticated approach of Hasenbusch
et al.~\cite{hasen_fp} to extend the Monte Carlo data to larger
lattice sizes.

\begin{figure}[htbp]
\includegraphics*[width=9.5 cm]{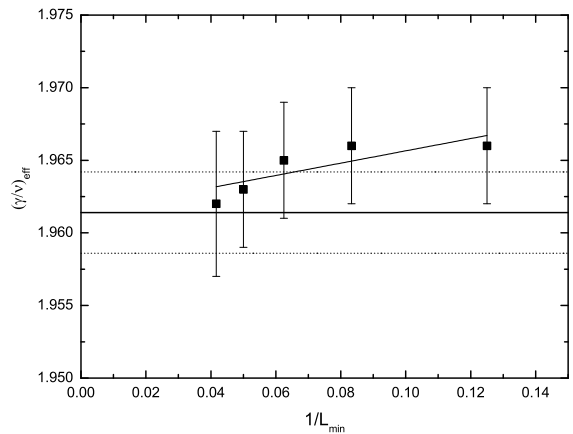}
\caption{\label{fig:4} (color online) Illustration of the behavior
of the effective exponents $\gamma / \nu$. The solid line drawn in
the panel, together with the dot lines, indicate the
unsatisfactorily estimated critical exponent range $\gamma /
\nu=1.9614(28)$.}
\end{figure}

We proceed to calculate the critical exponent ratio $\gamma/\nu$
from the peaks of the sample-averaged susceptibility
($[\chi]_{av}^{\ast}$). We assume that these finite-size anomalies
obeys a simple power law: $[\chi]_{av}^{\ast}=b\cdot
L^{\gamma/\nu}$, and follow again the practice of observing the
behavior of effective exponents by varying the $L_{min}$ of the
fitting range. The resulting sequence of effective exponents
($L_{min}=[8-24]$) is illustrated in Fig.~\ref{fig:4}. As seen
from this figure, the behavior of these estimates is not linear.
Thus, the attempted linear fit, appears as an overestimation, but
gives clearly the underestimated estimate $\gamma/\nu=1.9614(28)$,
appreciably smaller than that of Ballesteros \emph{et al.}
\cite{balles_rdi} $\gamma/\nu=1.963(5)$ and even more than those
of Hasenbusch \emph{et al.} $1.964(1)$ \cite{hasen_rdi} for the
site-diluted 3d Ising model and $1.964(2)$ \cite{hasen_fp} for the
corresponding to the present model isotropic $\pm J$ Ising model
at the ferromagnetic-paramagnetic transition line. Therefore, the
estimation via the susceptibility peaks is here unsatisfactory.

\begin{figure}[htbp]
\includegraphics*[width=9.5 cm]{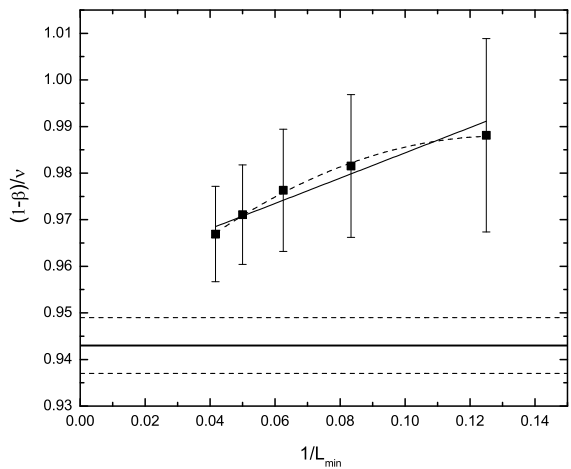}
\caption{\label{fig:5} (color online) Illustration of the behavior
of the effective exponents $(1-\beta)/ \nu$. The solid and the
dash lines drawn through the points indicate linear and
second-order polynomial fits, discussed in the text. The critical
exponent range for the later fit is indicated by the heavy solid
and dot lines at $(1-\beta)/ \nu=0.943(6)$.}
\end{figure}

We may now attempt the estimation of the exponent ratio
$\beta/\nu$, via the scaling behavior of the peaks corresponding
to the absolute order-parameter derivative which is expected to
scale as $[\partial \langle |M|\rangle /
\partial K]_{av}^{\ast}=b\cdot L^{(1-\beta)/\nu}$.
The corresponding effective exponent estimates, illustrated in
Fig.~\ref{fig:5}, show again a behavior which, as seen, is not
linear. The illustrated in the figure linear fit gives
${(1-\beta)/\nu}=0.9572(54)$, which by using our estimate $1 /
\nu=1.463(3)$, produces for $\beta/\nu$ the range $0.5058(84)$.
However, this value is noticeably smaller than the value
$\beta/\nu=0.518$, expected from hyperscaling and accepting an
estimate for $\gamma / \nu$ of the order of the above mentioned
literature estimates (say for instance: $\gamma / \nu=1.964$). The
situation in this case can be improved by attempting the linear
fit in Fig.~\ref{fig:5} only at the last three points. The linear
fit in these three points ($L_{min}=16,20,24$), gives
${(1-\beta)/\nu}=0.948(3)$ producing, as above,
$\beta/\nu=0.515(6)$. Note also, that the second order polynomial
fit, shown also in Fig.~\ref{fig:5} and applied to all five
points, gives an estimate ${(1-\beta)/\nu}=0.943(6)$ producing now
$\beta/\nu=0.520(9)$. These values are in good accordance with
hyperscaling and the literature estimate of $\gamma / \nu$.

The critical temperature will be now estimated by a simultaneous
fitting approach, using several pseudocritical temperatures of the
sample average of the quantities
measured~\cite{malakis09,malakis12}, as outlined earlier. The
simultaneous fitting is attempted to the expected power-law shift
behavior $T_{[Z]_{av}^{\ast}}=T_{c}+b_{Z}\cdot L^{-1/\nu}$ for the
six pseudocritical temperatures mentioned in the previous section.
\begin{figure}[htbp]
\includegraphics*[width=9.5 cm]{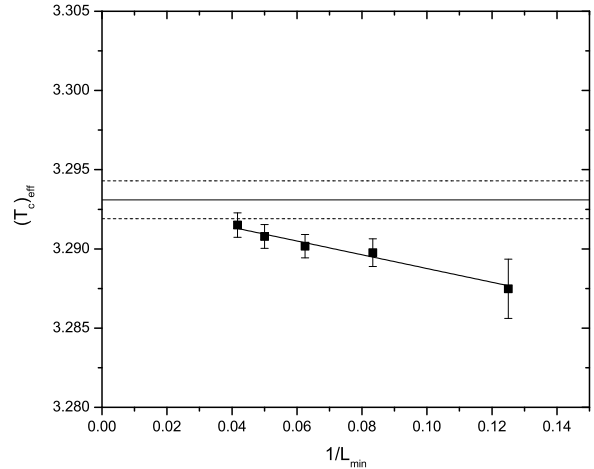}
\caption{\label{fig:6} (color online) Illustration of the behavior
of the effective pseudocritical temperatures obtained by applying
a simultaneous fit on the shift behavior $T_{[Z]^*}=T_c +b_Z
L^{-1/\nu}$ and fixing the exponents to the accurate estimate $1/
\nu=1.463$. The solid dot lines indicate the critical temperature
range $T_c=3.2931(12)$ obtained by the illustrated linear fit.}
\end{figure}

We approach this estimation by simultaneous fittings in which we
are fixing the exponent $1/\nu$ to the apparently accurate
estimate $1/\nu=1.463$. Following our earlier practice of using
different fitting ranges by varying the $L_{min}$ of the fitting
range, we obtain a sequence of estimates illustrated in
Fig.\ref{fig:6}. The linear fit shown in the panel gives an
estimate $T_c=3.2931(12)$ illustrated with solid and dot lines in
this figure, whereas restricting the fit only to the last three
points, corresponding to $L_{min}=16,20,24$, gives a higher
estimate $T_c=3.2945(18)$. We note here that a completely free
fit, without fixing any parameter, and following the above
practice gives $T_c=3.2934(8)$ from the linear fit to the five
points $L_{min}=8,12,16,20,24$ and $T_c=3.2940(16)$ from the
linear fit to the three points $L_{min}=16,20,24$. Thus, we will
suggest that $T_c=3.2938(9)$

\begin{figure}[htbp]
\includegraphics*[width=7.5 cm]{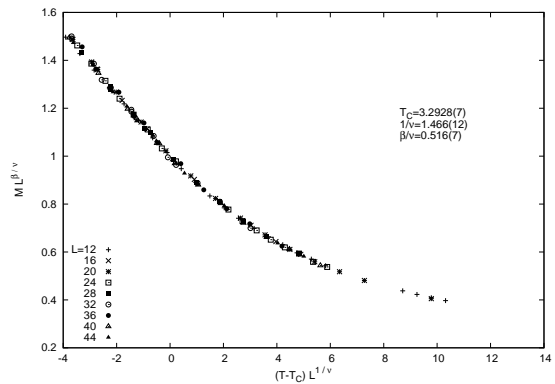}
\caption{\label{fig:7} (color online) Illustration of
magnetization data collapse for lattice sizes $L={12-44}$. The
method used is discussed in detail in the text and gives as shown
in the panel an optimum collapse at the values $T_c=3.2928(7)$, $1
/ \nu=1.466(12)$, and $\beta / \nu = 0.516(7)$.}
\end{figure}

Finally, we present the alternative estimation of critical
behavior by studying the expected scaling law Eq.~(\ref{eq:7}) for
the order parameter data, here the disorder averaged magnetization
data. Using the earlier mentioned collapse method, we rescale the
$y$ axis to $ML^{\beta / \nu}$ and the $x$ to $(T-T_c)L^{1/ \nu}$,
and we attempt to observe the optimum collapse of the
magnetization curves taken from different lattice sizes
($L=[12-44]$). We apply the downhill simplex algorithm as
developed and implemented in reference~\cite{collapse} for the
estimation of critical properties and their error bounds. As shown
in the panel of Fig.\ref{fig:7}, the optimum collapse gives
$\beta/ \nu=0.516(7)$. This value is in very good agreement with
the expected value as mentioned earlier. In the panel we also
give, the resulting estimates for the exponent $1/ \nu=1.466(12)$
and the estimate for the critical temperature $T_c=3.2928(7)$.
These are in fair agreement with our previous findings. However,
the estimate $\beta/ \nu=0.516(7)$, appears to be very
satisfactory.

\section{Conclusions}
\label{sec:4}

The present paper pointed out clearly that, the $\pm J$
three-dimensional Ising model, with spatially uniaxially
anisotropic bond randomness, give rise to a second-order phase
transition belonging in the same universality class with the 3d
random Ising model. The implemented anisotropy appears as an
irrelevant parameter for the ferromagnetic-paramagnetic transition
line. We found the reliable estimates $\beta/ \nu=0.516(7)$, by
using the collapse method of reference~\cite{collapse} and
$\nu=0.6835(25)$, from the smooth behavior of the logarithmic
derivatives of the order parameter. We have presented also a
conjectured global phase diagram, providing interesting
predictions.

Currently, we are carrying out further numerical simulations. From
these, it seems that, the implemented here anisotropy ($p_z=0 ;
J_{ij}^z=J_{ij}^{xy}$) is also an irrelevant parameter for the
other transition lines of the phase diagram (the
ferromagnetic-spin glass and the spin glass-paramagnetic lines).
Finally, we are considering the more general case ($J_{ij}^z \neq
J_{ij}^{xy}$). We hope that, we will soon provide further
confirmation of the discussed in this paper predictions and
observe and verify the interesting features, of the global phase
diagrams, brought out by the study of Guven et al.~\cite{anis_1}.

\begin{acknowledgments}
The authors are grateful to A. N. Berker for our constructive
discussions on the subject. This work was supported by the special
Account for Research of the University of Athens (code: 11112).
T.P has been supported by a Ph.D grand of the Special Account of
the University of Athens.
\end{acknowledgments}

\end{document}